\documentclass[prl,superscriptaddress,twocolumn,showpacs,preprintnumbers,amsmath,amssymb]{revtex4}

\usepackage{graphicx}
\usepackage{epstopdf}
\usepackage{dcolumn}
\usepackage{bm}
\usepackage{color}

\usepackage{times}
\usepackage{amsmath}
\usepackage{amssymb}
\usepackage{amsfonts}
\usepackage{latexsym}
\usepackage{verbatim}

\usepackage{color}

\usepackage[colorlinks,bookmarks=false,citecolor=blue,linkcolor=red,urlcolor=blue]{hyperref}

\def\vec{\mathbf}

\newcommand{\bea}{\begin{eqnarray}}
\newcommand{\eea}{\end{eqnarray}}
\newcommand{\be}{\begin{equation}}
\newcommand{\ee}{\end{equation}}

\newcommand{\la}{\langle}
\newcommand{\ra}{\rangle}

\newcommand{\K}{\mathbf{k}}

\newcommand{\si}{\mathbf{S}_i}

\newcommand{\sj}{\mathbf{S}_j}

\newcommand{\SCBO}{SrCu$_2$(BO$_3$)$_2$ }
\newcommand{\et}{{\it et al. }}

\begin{document}
\title{Emergence of One-Dimensional Physics in the Distorted Shastry-Sutherland Lattice}

\author{M.\ Moliner}\email{Marion.Moliner@kit.edu}
\affiliation{Institut f\"ur Nanotechnologie, Karlsruhe Institute of Technology, 76344 Eggenstein-Leopoldshafen, Germany.}

\author{I.\ Rousochatzakis}\email{rousocha@pks.mpg.de}
\affiliation{Max-Planck-Institut f\"ur Physik komplexer Systeme, N\"otnitzer Stra\ss e 38, 01187 Dresden, Germany.}

\author{F.\ Mila}\email{Frederic.Mila@epfl.ch}
\affiliation{Institute of Theoretical Physics, Ecole Polytechnique F\'ed\'erale de Lausanne, CH-1015 Lausanne, Switzerland.}

\date{\today}

\begin{abstract}
Motivated by the on-going investigation of SrCu$_2$(BO$_3$)$_2$ under pressure, 
we study a variant of the two-dimensional Shastry-Sutherland (SS) spin-$1/2$ model with two types of dimers.
Combined with the frustration of the SS model, this modification induces, in a large parameter range, a dimensional reduction at low energies, with nearly decoupled effective $S=1$ Haldane chains 
forming along one of the diagonals of the lattice. We also present evidence that the intermediate plaquette solid phase 
of the undistorted SS model remains stable in a finite region of the phase diagram.
\end{abstract}




\maketitle

\begin{figure}[!b]
\includegraphics[width=0.34\textwidth]{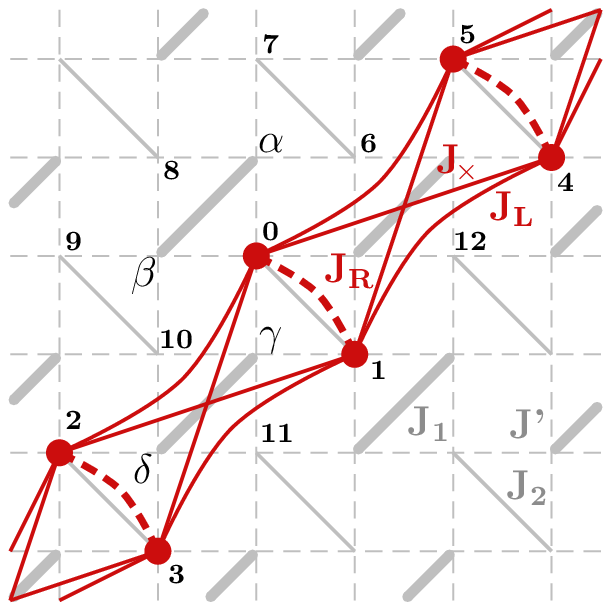}
\vspace{-2mm}
\caption{\label{fig:Lattices} (color online)
The distorted SS model (thick and thin gray lines for $J_1$, $J_2$, and dashed for $J'$) of this work,
together with the effective quasi-1D model (red solid and dashed thick lines) emerging in the limit $J_1 \gg J_2, J'$ which gives rise to the Haldane phase.
}
\end{figure}

\textit{Introduction}.---
An exciting route for exploring new phases in frustrated magnets consists in applying hydrostatic pressure so as to considerably modify the dominant magnetic interactions 
and thus alter the relevant low-energy degrees of freedom. The quasi-2D compound \SCBO \cite{review_SCBO} offers such a possibility
since it is in the close vicinity of a quantum phase transition from an exact tensor product of singlets
toward a plaquette solid\cite{Koga_2000,Laeuchli_2002}.
Indeed, Nuclear Magnetic Resonance (NMR)\cite{Waki_2007,Takigawa_2010}, Inelastic Neutron Scattering (INS)\cite{Zayed},
as well as earlier susceptibility measurements\cite{Kageyama_2003}, show strong evidence that \SCBO undergoes at least one phase transition under high hydrostatic pressure, the nature of which is still not fully understood. 
What is clear however is that the four-fold (C$_4$) symmetry around the void plaquettes
is quickly lost under pressure, so that nearest-neighbor dimers pointing in different directions become
inequivalent (see Fig.\ref{fig:Lattices}). This naturally leads to an extension of the Shastry-Sutherland (SS) model with two different diagonal bonds $J_1$ and $J_2$ described by the spin $S=1/2$ Heisenberg Hamiltonian
\be
\mathcal{H} =  J' \sum_{\la ij \ra}  \si \cdot \sj
                    + J_1\!\!\sum_{\ll ij \gg_1}\!\!\!\! \si \cdot \sj
                    + J_2 \!\!\sum_{\ll ij \gg_2}\!\!\!\! \si \cdot \sj
\label{Hamiltonian}
\ee
where $J'$ denotes the nearest-neighbor exchange (see Fig. \ref{fig:Lattices}).
Since all nearest-neighbor bonds $J'$ are equal, the product of singlets on $J_1$ and $J_2$
bonds is always an eigenstate, as it is for the undistorted SS model. Besides, although the C$_4$ symmetry is lost, 
both the sets of $J_1$ and $J_2$ bonds form a square lattice.
It thus seems natural to expect that the phase diagram will consist of three phases, as for the regular SS model:
the exact singlet phase for small $J'$, the N\'eel phase for large $J'$, and an intermediate plaquette phase with 2D character\cite{Koga_2000,Laeuchli_2002}.

\begin{figure}[!t]
\includegraphics[width=0.48\textwidth]{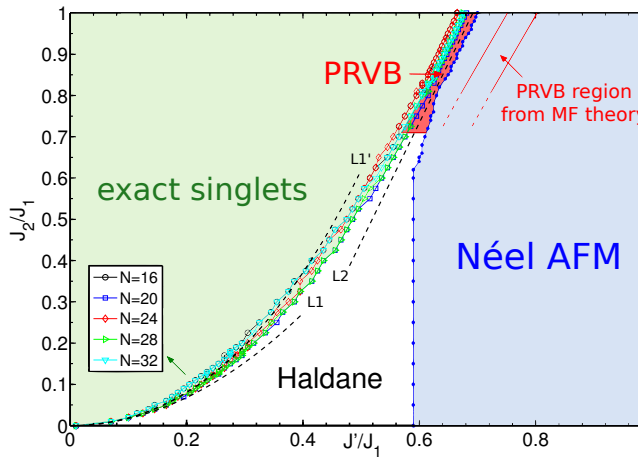}
\caption{\label{fig:Phase_Diagram} (color online) The phase diagram of the present model as obtained by exact diagonalizations (symbols with lines).
The red lines enclose the stability region of the PRVB phase as obtained from the quadrumer boson operator method (see text).
The dashed lines L1 
and L1' 
denote the boundary between the exact dimer and the Haldane phase as obtained from perturbation theory up to second and third orders, 
while the dashed line L2 
is the line along which we examine the low-E spectrum shown in Fig. \ref{fig:LESpectrum_dline} below.
}
\end{figure}

In this paper, using a variety of analytical and numerical techniques, we show that making the diagonal bonds
inequivalent actually triggers an abrupt dimensional reduction into a new, effectively 1D Haldane phase
in a wide parameter range (see Fig. \ref{fig:Phase_Diagram}).
In this phase the effective low-energy physics can be described in terms of nearly decoupled, fully frustrated 2-leg ladders (see Fig. \ref{fig:Lattices})
running along one of the two diagonals of the square lattice. Remarkably enough, the spins on the $J_2$-bonds form {\it triplets} in this phase, leading to a completely different
picture of the low-lying magnetic excitations as compared to the exact singlet phase.


The emergence of this new singlet phase is best understood by starting from the limit $J_2/J_1=J'/J_1=0$.
The model then reduces to a set of isolated dimers on the $J_1$-bonds and isolated free spins on the remaining sites (see Fig.~\ref{fig:Lattices}), resulting into a macroscopically degenerate ground state manifold. The degeneracy is lifted by switching on the inter-dimer coupling $J'$ which, to leading (second) order, {\it a priori} couples a given spin to
12 neighbors (see Fig.~\ref{fig:Lattices}, where the 12 neighbors of site 0 are numbered from 1 to 12). Each second-order process involves an intermediate
state with a $J_1$ dimer in a triplet state, and the resulting coupling is ferromagnetic and of amplitude $-J'^2/2J_1$
if the $J'$ bonds are connected to the same site of the $J_1$ dimer, while it is antiferromagnetic and of amplitude $J'^2/2J_1$ if they are connected to opposite sites.
When processes with different signs couple two spins, they cancel out due to quantum interferences. As a result, the couplings between site 0 and
sites 6 to 12 vanish, and site 0 is only coupled to its neighbors on the ladder shown in Fig.~\ref{fig:Lattices}.
The couplings between site 0 and sites 2, 3, 4 and 5 are all antiferromagnetic and of amplitude $J'^2/2J_1$, while the coupling between 0 and 1 is ferromagnetic and equal to $-J'^2/J_1$.

Including both $J_2$ and $J'$, and up to second order in $J'/J_1$, our 2D model thus reduces to an infinite set of 2-leg,
frustrated spin ladders directed along the $J_1$ diagonal bonds. The effective interactions within the ladders are depicted in Fig. \ref{fig:Lattices} on top of the original model.
They consist of the rung coupling $J_R=J_2-J'^2/J_1$, the leg coupling $J_L=J'^2/2J_1$, and a frustrating coupling $J_{\!\times}=J_L$.
This model has been studied quite extensively\cite{Honecker_2000,IvanovRichter}. It has the special property that the total spin on each rung
is a good quantum number, which allows to determine a large number of eigenstates exactly\cite{Honecker_2000}.
In particular, in the limit $J_R \gg J_L$ (or $J_2 \gg J'^2/J_1$) the lowest energy is obtained by minimizing the number of triplets in the rungs\cite{Bose_1993}.
The resulting ground state of the ladder is the product of rung singlets, which together with the strong $J_1$ singlets correspond to the exact orthogonal-dimer phase of the SS model.
On the contrary, if  $J_R$ is not too large (and {\it a fortiori} if it is negative), fluctuations between neighboring triplets dominate and the lowest energy is obtained by placing a triplet on each rung.
The ladder then behaves like a spin-1 chain, and hence the ground state corresponds to the Haldane gapped phase\cite{Haldane_1983}.
The actual transition between the exact dimer and the Haldane phase is known\cite{Honecker_2000} to take place at $J_R/J_L \simeq 1.4$ and it is of first order.
Rewritten in terms of the original couplings of our model, this boundary corresponds to the line $J_2 = 1.7\, J'^2/J_1$, which is shown by the dashed line L1 in Fig. \ref{fig:Phase_Diagram}.
The dashed line L'1, defined by $J_2=1.7 J'^2/J_1 + 1.55 J'^3/J_1^2$, corresponds to the boundary obtained from perturbation theory up to third
order\footnote{Up to third order, the effective ladder couplings are $J_R=J_2-J'^2/J_1 - \frac{1}{2} J'^3/J_1^2$ and $J_L=J_{\times}= \frac{1}{2} J'^2/J_1 + \frac{3}{4} J'^3/J_1^2$.}.
Third order processes in $J'/J_1$ also induce some coupling between nearest and next-to-nearest neighboring ladders, resulting into a
complex network\footnote{Site 0 couples to 6 and 10 through $J'^3/J_1^2$, and to 7, 9, 11 and 12 through $\frac{1}{2} J'^3/J_1^2$.
Site 0 is also coupled to site 8 belonging to the next-nearest-neighbor ladder through $J'^3/J_1^2$.}, but the quasi-1D character of the (short-range) spin correlations and the finite spin gap seem to be a robust feature of this phase, as suggested by exact diagonalizations. 

\begin{figure}[!t]
\includegraphics[width=0.45\textwidth]{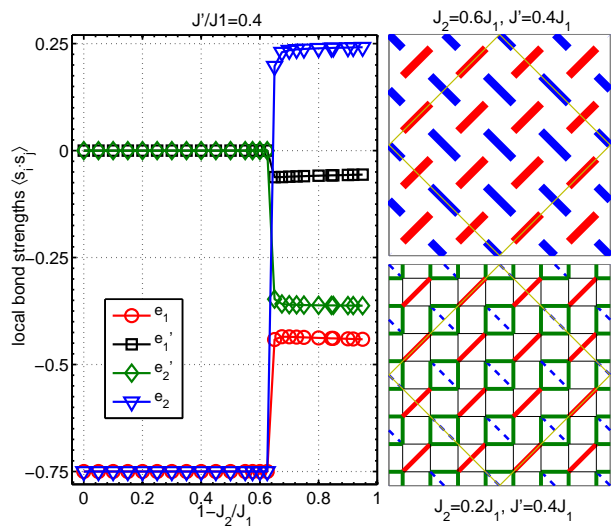}
\caption{\label{fig:BondProfiles} (color online) Left panel: Ground state expectation values
of the four different nearest-neighbor exchange energies $\vec{S}_i\cdot\vec{S}_j$ for the 32-site cluster
($e_{1,2}$ for the $J_{1,2}$ diagonal bonds, and $e_{1,2}'$ for $J'$-bonds on plaquettes which contain $J_{1,2}$ bonds).
Right panels: The corresponding local energy profiles for representative cases inside the exact singlet state (above) and the Haldane state (below).
Positive (negative) exchange values are shown by dashed (solid) lines, whose width is proportional to the magnitude of these values. 
}
\end{figure}

\textit{Exact diagonalizations}.---
To determine the behavior of the present model away from the above perturbative regime and to accurately obtain the full phase diagram of the model,
we have performed an extensive exact diagonalization study (ED) on finite-size clusters with periodic boundary conditions.
Apart from $\mathsf{U}(1)$ spin rotations and translational invariance we have also exploited the two reflection symmetries (denoted here by $\mathcal{R}_1$ and $\mathcal{R}_2$)
along the two diagonals of the square lattice.

The transition between the exact singlet phase and the Haldane phase is most simply identified
by measuring the ground state spin-spin correlations $\la \vec{S}_i\!\cdot\!\vec{S}_j\ra$ on neighboring bonds, as shown in Fig.\ref{fig:BondProfiles}: they are rigourously equal to $-3/4$ on dimer bonds and to zero on $J'$ bonds in the exact dimer phase, and jump upon entering the Haldane phase to ferromagnetic and nearly equal to $1/4$ on the $J_2$ bonds, and to intermediate antiferromagnetic values on $J_1$ and $J'$ bonds. The resulting first order transition lines shown in Fig. \ref{fig:Phase_Diagram} are in good agreement with the perturbative predictions and show a weak system size dependence. In addition, the correlation profiles shown in the right panels of Fig. \ref{fig:BondProfiles} demonstrate how the 2D-lattice decouples into almost non-interacting 2-leg ladders.

\begin{figure}[!t]
\includegraphics[width=0.44\textwidth]{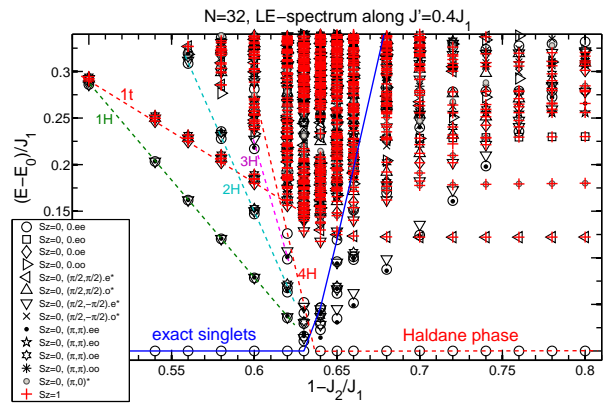}
\caption{\label{fig:LESpectrum_vline}
(color online) Low-energy spectrum (measured from the GS) for the 32-site cluster along the line $J'/J_1=0.4$ in the $S_z=0$ and $1$ sectors.
The blue solid line shows the energy of the exact singlet phase.
The dashed lines highlight the n-Haldane chain singlets discussed in the text.
The quantum numbers shown in the legends correspond to the allowed momenta of the 32-site cluster and the parities with respect to the two diagonal reflections $\mathcal{R}_{1,2}$.
}
\end{figure}

Next, we examine the behavior of the spin gap by looking at the evolution of the low-energy excitation spectrum as we move away from the exact dimer into the Haldane phase.
Figure \ref{fig:LESpectrum_vline} shows the low-energy spectrum of the 32-site cluster along the line $J'=0.4J_1$ in the $S_z=0$ and $1$ sectors.
All energies are measured from the ground state.
In the exact dimer phase the spin gap scales linearly with $J_2$ and thus decreases as we go towards the Haldane phase.
This happens because a magnetic $S=1$ excitation results by promoting a rung singlet\footnote{We note here that a triplon excitation prefers to sit on the $J_2$-bonds since $J_2<J_1$.}
into a triplet which costs energy $J_2$ (disregarding the polarization corrections to the spin gap driven by $J'$\cite{review_SCBO}).
By contrast when entering the Haldane phase at $J_2\simeq 0.365 J_1$ the gap does not change with $J_2$ any longer.
This happens because once we pay the energy to form the effective spin-1 entities on the rungs the energy scale $J_2$ does not enter their effective dynamics any longer.

The 1D character of the Haldane phase underlies also a number of striking spectral features in the singlet sector.
In particular, Fig. \ref{fig:LESpectrum_vline} shows a number of singlets (highlighted by dashed lines) which group together into almost degenerate levels, and all seem to become soft together with the Haldane state (red dashed line). These correspond to singlets where not all ladders are in the Haldane phase.
Specifically, the 32-site cluster (enclosed by solid yellow lines in the right panels of Fig. \ref{fig:BondProfiles}) decomposes into four 2-leg ladders with two $J_2$-bonds each.
To form the Haldane phase in a single ladder we pay energy $2 J_2$ to create two triplets (disregarding again the polarization energy driven by $J'$), but we gain an energy $\delta_H$ from their interaction.  This gives 4 possible states with energy $E_{1H}\simeq 2J_2-\delta_H$ whereby one ladder has entered the Haldane phase but the remaining three have still singlets in their rungs.
Similarly, there must exist: 6 states with two ladders in the Haldane phase and energy $E_{2H} \simeq 2E_{1H}$;
4 states with three ladders in the Haldane phase and $E_{3H} \simeq 3E_{1H}$;
and finally a single state where all ladders are in the Haldane phase with $E_{4H} \simeq 4E_{1H}$.
Both the multiplicities and the dependence on $J_2$ (the slope in the dashed lines of Fig.  \ref{fig:LESpectrum_vline}) are in agreement with the ED spectra.
In addition, we have derived the decomposition of each group of singlets into irreducible representations of the space group of the 32-site cluster and the results also agree with the ED spectra. Finally, the fact that $E_{nH} \simeq n E_{1H}$ explains why the above groups of singlets seem to become soft together with the Haldane state.

We note here that the above energy considerations are only true to leading order in $J'$. There are various perturbative corrections
which renormalize the energy of each of the above singlets in a different manner. For the 32-site cluster, it is the global Haldane state that crosses the exact singlet phase first.

\begin{figure}
\includegraphics[width=0.45\textwidth]{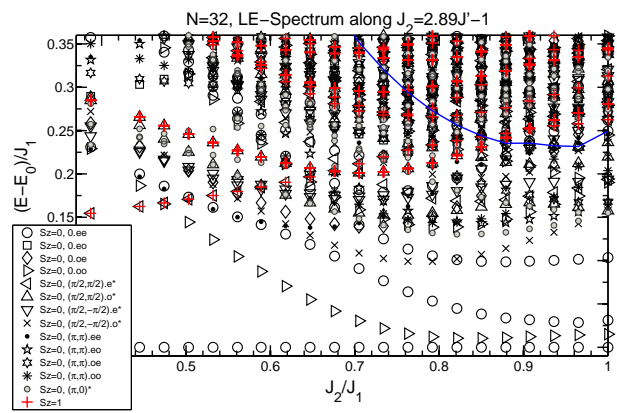}
\caption{\label{fig:LESpectrum_dline}
(color online) Low-energy spectrum (measured from the GS) of the 32-site cluster along the line $L_2$  ($J2=2.89J'-1$) of Fig. \ref{fig:Phase_Diagram} in the $Sz=0$ and $1$ sectors.
The solid (blue) line stands for the energy of the exact singlet phase.
}
\end{figure}

Upon increasing $J'/J_1$, the Haldane phase undergoes a second order transition into the N\'eel phase.
The disappearance of the N\'eel order is continuous and corresponds to the vanishing of the 
local moment $m$ obtained by a usual\cite{Laeuchli_2002} finite-size extrapolation $m_N \simeq m+\frac{c}{\sqrt{N}}$,
using numerical data for the corresponding spin structure factor at zero momentum.\footnote{For the N\'eel boundary $J'_{\text{cr}}(J_2)$ of Fig. \ref{fig:Phase_Diagram} we made an unbiased use 
from all clusters with 16, 20, 24, 28, and 32 sites. Using the most symmetric clusters (16 and 32 sites) shifts $J'_{\text{cr}}$ by no more than a few percent.}  
We find that for $J_2\lesssim 0.6 J_1$,  the N\'eel boundary does not depend on $J_2$, which is related to the absence of this energy scale from the effective dynamics of the Haldane phase.
On the other hand, the bending of the N\'eel boundary for $J_2 \gtrsim 0.6 J_1$ indicates the proximity to another phase, different from the Haldane one.
We now show evidence, first by a detailed spectral analysis and then by a perturbative mean-field approach, that this phase is adiabatically connected to the plaquette  phase present for $J_2=J_1$\cite{Koga_2000,Laeuchli_2002}. 
We begin by examining the evolution of the low-lying spectrum along the line L2 of Fig. \ref{fig:Phase_Diagram}.
As shown in Fig. \ref{fig:LESpectrum_dline} for the 32-site spectrum, a low-energy excitation becomes gradually soft above $J_2/J_1\sim 0.7-0.8$. This excitation has zero momentum 
but it has odd parity with respect to $\mathcal{R}_1$ and $\mathcal{R}_2$, in contrast to the ground state.
If these two states become degenerate in the thermodynamic limit then the above quantum numbers are exactly the ones expected for the $\mathsf{Z}_2$ plaquette phase,
since the latter does not break translational invariance but it does break $\mathcal{R}_1$ and $\mathcal{R}_2$. 

We also find a large number of singlets below the lowest triplet excitation. The majority of these states have a positive curvature with $J_2$, similarly to the
exact singlet state shown by the solid (blue) line. Hence these states are most probably adiabatically connected to
singlets above the dimer phase, such as the ones we discussed above that must become nearly soft at the boundary with the Haldane phase
(in addition there could also be some bound states of triplons\cite{Knetter_2000}).
In agreement with L\"auchli \et\cite{Laeuchli_2002}, we also find a singlet which lies slightly above the lowest two singlets at $J_2=J_1$.
The nature of this singlet might well be a singlet excitation (e.g. a domain wall) of the plaquette phase.

A complementary confirmation that the plaquette phase survives in a small region for $J_2\neq J_1$ can be also given by 
an analytical mean-field approach based on quadrumer boson operators\cite{Bhatt_1990, Zhitomirsky_1996,Laeuchli_2002,Starykh_2005}.
Here, one essentially expands around the explicitly symmetry broken phase where every second void square is in the 
singlet ground state of an isolated Heisenberg plaquette\cite{Koga_2000}. The ground state and low-lying excitations of the original model are then approximated by truncating
the Hilbert space of each plaquette to its singlet ground state and its lowest triplet, which is achieved by the bosonic representation 
of the four spins ($j=1,\ldots 4$)
\be
S_{j}^{\alpha} = \frac{(-1)^j}{\sqrt{6}}\big( s^{\dagger} t_{\alpha} + t_{\alpha}^{\dagger} s \big)-\frac{i}{4}\sum_{\beta,\gamma}\epsilon^{\alpha\beta\gamma}t_{\beta}^{\dagger}t_{\gamma}, 
\ee
where $\alpha=x, y, z$, and $s^{\dagger}$, $t_\alpha^{\dagger}$ create respectively the singlet ground state and the lowest triplet excitation of a single Heisenberg  plaquette.
A hard-core constraint restricts the boson occupation number of each plaquette $s_{}^{\dagger}s_{}^{} + \sum_{\alpha}t_{\alpha}^{\dagger}t_{\alpha}^{}=1$ so that the bosonic representation obeys the spin algebra. The energy of a single plaquette becomes $\mathcal{H}_P=E_s s^{\dagger}s + E_t  \sum_\alpha t_{\alpha}^{\dagger}t_{\alpha}$.
Following the standard procedure\cite{Laeuchli_2002, Zhitomirsky_1996,Starykh_2005}, we obtain a triplet mode with dispersion (in units of $J_1=1$)
\be
\omega^2(\K)=J'^2+\frac{2}{3} J'\Big[ (1 - 2 J')\cos k_x +  (J_2 - 2 J') \cos k_y \Big]
\label{eq:Quad_bosons_dispersion}
\ee
For $J'>\frac{1}{2}J_1$ and $J_2<2J'$ the minimum of the spectrum is at $\K=0$ and this becomes soft along the line $J_2=-1 + \frac{5}{2}J'$,
which corresponds to the N\'eel ordering.
The first-order transition line between the dimer and the plaquette phase is obtained by comparing the exact dimer energy
with that of the plaquette state including the zero-point motion from the quadratic terms.
These lines are depicted as red lines in Fig. \ref{fig:Phase_Diagram}. The stability region is shifted
to larger $J'/J$, as already observed for the standard SS model, but the slopes
are in good agreement with ED. Note that the first order boundary between the plaquette and the
Haldane phase cannot be found with our method since we do not have an expression for the ground state energy of the Haldane phase.

\textit{Conclusion}.---
We have introduced a variant of the SS model with two types of diagonal couplings as a minimal model for the loss of the C$_4$ symmetry in \SCBO under high 
pressure. 
We find a novel quasi-1D Haldane phase where the low-energy physics can be described in terms of fully frustrated decoupled 2-leg ladders which form $S=1$ entities on their rungs.
When $J_2/J_1$ decreases, the gap to the first excitation
first decreases linearly, and then levels off after crossing
the boundary between the exact dimer phase and the Haldane
phase. Assuming that the main effect of pressure is to reduce the ratio $J_2/J_1$,
this is precisely what has been reported by INS\cite{Zayed}, as well as
by NMR\cite{Takigawa} for the nonmagnetic sites. However, the presence of magnetic
sites possibly arranged on every second ladder, as suggested by NMR\cite{Waki_2007},
and of a low-lying excitation below the excitation that levels off, as revealed
by INS\cite{Zayed}, goes beyond the prediction of the present model and requires
further investigation.

\begin{acknowledgments}
We would like to thank A.  L\"auchli, H. M. R\o nnow, M. Zayed, C. Ruegg, M. Takigawa, S. Wessel and K. Penc for fruitful discussions. This work has been supported by the Swiss National Fund and by MaNEP.
\end{acknowledgments}

\end{document}